\documentstyle[bibstyle]{article}

\newcommand{\tabref}[1]{Table~\ref{#1}}
\newcommand{\figref}[1]{Figure~\ref{#1}}
\newcommand{\secref}[1]{Section~\ref{#1}}

\newcommand{\Figref}[1]{\figref{#1}}

\newcommand{\coude}{coud\'{e}}
\newcommand{\dSco}{\mbox{$\delta$~Sco}}

\newcommand{\about}{\mbox{$\sim$\,}}
\newcommand{\degree}{\mbox{$^\circ$}}
\def\sol{\mbox{$_\odot$}}
\def\farcs{\hbox{$.\!\!^{\prime\prime}$}}
\def\micron{\hbox{$\mu$m}}
\def\deg{\hbox{$^\circ$}}

\newcommand{\mycaption}[3]
{\if*#2 \caption{#3\label{#1}}
 \else  \caption[#2]{#3\label{#1}}
 \fi}

\input{psfig}

\title{The Orbit of the Binary Star Delta Scorpii}

\author{Timothy R. Bedding\\[1ex]
School of Physics, University of Sydney, 2006, Australia
\\
and
\\
European Southern Observatory, Karl-Schwarzschild-Stra\ss e 2,
\\
Garching bei M\"unchen, D-8046 Germany\thanks{current address}
\\
E-mail: tbedding@hq.eso.org}

\date{Received:18 February 1993; Accepted: 26 April 1993}

\begin{document}

\maketitle

\begin{abstract}
Although \dSco\ is a bright and well-studied star, the details of its
multiplicity have remained unclear.  Here we present the first
diffraction-limited image of this 0\farcs12 binary star, made using optical
interferometry, and resolve the confusion that has existed in the
literature over its multiplicity.

Examining published speckle measurements, together with the present result,
reveals a periodicity of 10.5\,yr and allows calculation of the orbital
parameters.  The orbit has a high eccentricity ($e=0.82$) and large
inclination ($i=70\deg$), making it a favourable target for radial velocity
measurements during the next periastron (in 2000).
\end{abstract}

\section{Introduction}

The star \dSco\ (HR~5953) is known to be a multiple system, although there
has been some confusion in the literature over the exact degree of
multiplicity.  Its entry in the {\em Bright Star Catalogue\/}
\cite{Hoffleit} indicates that the primary has companions at separations of
0\farcs1 and 0\farcs18.  Close triple stars are rare and make excellent
targets for optical interferometry.  An observation of \dSco\ made with
MAPPIT, an optical interferometer at the Anglo-Australian Telescope, is
presented in \secref{sec.obs}.  The resulting image reveals only two
components, which prompted me to search the literature for clarification.
The origin of the confusion makes an interesting tale and appears in
\secref{sec.hist}.

In addition, inspection of published speckle observations (together with
the MAPPIT measurement) reveals a periodicity of 10.5\,yr and allows
determination of the orbital elements of the system.  As shown in
\secref{sec.orbit}, radial velocity variations should be measurable near
periastron, making \dSco\ one of the brightest in a growing list of stars
which are both visual and spectroscopic binaries.


\section{Observations}	\label{sec.obs}

MAPPIT (Masked APerture-Plane Interference Telescope) is an optical
interferometer located at the \coude\ focus of the 3.9-m Anglo-Australian
Telescope \cite{BRM92,Bed92}.  The instrument uses the technique of
non-redundant masking \cite{HMT87,NKG89}, in which short-exposure images
are recorded through an aperture mask containing a small number of holes.
The images are analysed to determine the power spectrum and closure phases
of the object, which can be used to reconstruct a true diffraction-limited
image.

One advantage of using a non-redundant aperture mask is that it increases
the signal-to-noise ratios of the power spectrum and closure phase
measurements relative to observations with an unobstructed aperture.  This
is despite the fact that much of the light is blocked by the mask.  Another
advantage is that it improves the accuracy with which one can correct for
variations in atmospheric seeing, something which is often the limiting
factor in high-resolution imaging.  The main drawback of non-redundant
masking is a less efficient coverage of spatial frequencies.  However, for
simple objects such as multiple and barely resolved stars, adequate spatial
frequency coverage can be obtained by combining observations made with
different masks and with the masks rotated to several different position
angles on the sky.

The observations reported here were made on the nights of 31~May 1991 and
1~June 1991.  The detector was the Image Photon Counting System (IPCS;
{}\citebare{Bok90}), operated in high-speed mode with a video frame time of
6.5\,ms.  Two aperture masks were used, each with five 5-cm holes in a
linear non-redundant array.  The longest baselines on these masks were
2.2\,m and 3.3\,m, respectively.  We took seven sets of data, each
containing \about15000 frames and each made with the mask array set to a
different position angle on the sky.  Three of the data sets used the 3.3-m
mask and four used the 2.2-m mask.  The reconstructed image of {}\dSco\
shown in {}\figref{fig.dsco-image} was produced using standard
radio-astronomical methods (CLEAN and self-calibration).  More details of
the observations and the data processing methods may be found in
{}\citeone{MBR92} and {}\citeone{Bed92}.

\begin{figure}


\mycaption{fig.dsco-image}{Restored image of \dSco\relax}
{
Restored image of \dSco. Contour levels are at $-2$, 2, 5, 10, 20, 30, 50,
70 and 90\% of the peak; the dynamic range is about 50:1.  North is to the
top and east to the left.
}
\end{figure}

We find \dSco\ to be a double source with a separation of $0\farcs116\pm
0\farcs005$ at position angle $345\degree\pm5\degree$.  The use of closure
phases allows us to determine the orientation without the 180\degree\
ambiguity inherent in conventional autocorrelation processing.  The
magnitude difference is best extracted by fitting directly to the power
spectrum measurements, and we find a value of $\Delta m=1.5\pm0.3$.  There
is no evidence for a third component, either in the power spectra or in the
reconstructed image.

\section{Previous Studies of \dSco}	\label{sec.previous}

 \dSco\ has a combined spectral type of B0.3\,IV and a $V$ magnitude of
2.32 \cite{Hoffleit}.  It has been studied several times, both in its own
right and as a probe of the interstellar medium, and its multiplicity
therefore deserves clarification.  The following is a summary of some
previous observations of \dSco, including a discussion in \secref{sec.hist}
of the multiple nature of the star.

 \dSco\ has no record of photometric variability and so is not classified
as a $\beta$~Cephei star, although it does show line-profile variability
with a period of several hours.  This was discovered by \citeone{Smi86},
whose observations suggested a total of six non-radial modes of oscillation
and of `mode switching' on time scales of months.

The \dSco\ system lies in the $\rho$~Ophiuchi cloud, which is itself part
of the Upper Scorpius region of the Scorpio-Centaurus association.  The
$\rho$~Ophiuchi cloud is of interest as an area of high-density
interstellar dust, high depletions and low ultraviolet extinction
\cite{CSS73,S+J80}.  \dSco\ itself is surrounded by a roughly circular
patch of H$\alpha$ emission about 3\degree\ in diameter \cite{Siv74}.  This
emission appears to be associated with an expanding shell of neutral
hydrogen, possibly the remnants of a supernova explosion \cite{San74}.

The ultraviolet spectrum of \dSco\ is unusual in that it shows no evidence
for a strong wind \cite{S+M76}.  The star does have a high-velocity
outflow, but the mass-loss rate ($3\times10^{-11}$\,M\sol\,yr$^{-1}$) is
lower than for other stars of similar spectral class and luminosity
\cite{Sno81}.  In the infrared, a 60\,\micron\ {\em IRAS\/} image presented
by \citeone{vB+M88} shows an arc-like feature around \dSco, which these
authors interpret as thermal dust emission arising from a bow shock.  They
suggest that such bow shocks are formed by the interaction between the
interstellar medium and the stellar wind of a rapidly moving star.  In the
case of \dSco, it seems that such a mechanism is inconsistent with the low
mass-loss rate inferred from the ultra-violet spectra.  This may indicate
that the stellar wind has varied significantly.

Because of its brightness, particularly in the ultraviolet, \dSco\ has been
used in several studies of the interstellar medium.  Absorption-line
observations towards the star have been made in the sodium D-lines by
\citeone{Hob69} and in the lines of Mg\,{\sc i} and Mg\,{\sc ii} near
280\,nm using a balloon-borne spectrograph \cite{BCD76}.  These
observations revealed low-velocity absorption clouds comprising gas with
Mg/H and Na/H abundances approximately a factor of ten below solar.  There
was also a report by \citeone{HRR75} of two components of high-velocity
gas, at $-47$ and $-63$\,km\,s$^{-1}$, observed in the Na(D$_2$) line.
However, subsequent observations by \citeone{Hob76} failed to detect these
features and they were also absent in the ultraviolet Mg lines
\cite{BPB76}.  The consensus appears to be that the high-velocity
components seen by \citename{HRR75} were not interstellar sodium, and were
probably due to telluric absorption features.

More recently, \dSco\ was one of several stars used to determine extinction
in the far-ultraviolet caused by interstellar dust grains \cite{SAP90}.
The extinction curves, based on data from the two Voyager spacecraft, were
found to be consistent with theoretical predictions.  \citeone{SMV88} made
high-resolution sounding rocket observations of interstellar H$_2$ lines
towards \dSco.  They deduced that the ultraviolet radiation field in the
$\rho$~Oph region is about six times more intense than in the solar
neighbourhood.  Lower resolution spectra towards similar stars have shown
systematic velocity separations between lines arising from different
rotational levels \cite{Sp+M76}.  These results implied that the high-$J$
lines arose in material being expelled from the background O~or B~star.
For \dSco, \citeone{SMV88} found no such systematic velocity separations,
indicating that H$_2$ towards this star does not have a significant
component arising in an expanding circumstellar shell or bubble.  This is
consistent with \dSco\ having an usually low mass-loss rate.

\subsection{The Multiplicity of \dSco}	\label{sec.hist}

There is some evidence that \dSco\ is a single-lined spectroscopic binary
with a period of about a month and an amplitude of \about7\,km/s
{}\cite{vHBD63,LMM87}.  However, the inferred separation is unresolvable
with a 4\,m telescope and we therefore would not expect our MAPPIT
observation to detect the secondary, if it exists.

We now turn to the more distant component of the \dSco\ system, detected in
the MAPPIT image at a separation of~0\farcs12.  The `discovery' of this
component was published in 1974 in three independent papers:
\begin{enumerate}

 \item In the {\em Occultation Newsletter}, \citeone{Dun74} listed an
observation of a lunar occultation of \dSco\ made in April 1974 by
R.~Nolthenius.  The observation revealed a companion star at separation
0\farcs1 and position angle 335\degree.

 \item \citeone{LBS74} used speckle interferometry to find a `previously
unknown companion to \dSco.'  From observations made in February and May
1973, they measured a separation of 0\farcs18 at position angle 170\degree\
with $\Delta m=2$.

 \item Describing measurements made using the Intensity Interferometer,
\citeone{HDA74} wrote: `\dSco\ has not been listed previously as a multiple
star but the observed value of [the zero-baseline correlation] $C_N$
($0.75\pm0.07$) is consistent with a binary star with $\Delta m \simeq
1.9~[\pm0.5]$.'  These observations, which were made in April 1971, did not
provide information about the separation or position angle of the
components.

\end{enumerate}
There is a large discrepancy in position angles between the speckle and
occultation measurements.  We can account for this by recalling that the
speckle observations have a 180\degree\ ambiguity.  This now explains the
entry in the {\em Bright Star Catalogue\/}: the speckle and occultation
measurements are listed separately, although they actually describe the
same component.  From observations of a grazing occultation of \dSco,
\citeone{R+D88} confirmed that 180\degree\ needs to be added to the speckle
position angles (by that time, the star had been observed several times
using speckle interferometry).  The correct orientation is also confirmed
by the MAPPIT image.

Interestingly, it turns out that all three 1974 papers were wrong in their
claims to have discovered the companion to \dSco.  As \citename{R+D88}
({\em ibid}\/) point out, the star's duplicity was actually discovered
during a lunar occultation at the turn of the century when \citeone{Inn01}
observed that the star took several tenths of a second to disappear and
realized it was a double.

The {\em Bright Star Catalogue\/} also mentions an occultation companion at
a separation of 0\farcs00001, which must be an error.  It is certainly too
close to have been measured by lunar occultation techniques.  In any case,
the distance of \dSco\ (\about135\,pc; \citebare{dGdZL89}) would imply a
separation of only 0.3\,R\sol, whereas the radius of the primary star is
expected to be many times this.

\section{The Orbit of \dSco}	\label{sec.orbit}

Although \dSco\ has now been observed many times using speckle
interferometry, I have found no discussion in the literature of the orbit.
Indeed, it generally appears to be believed that the binary has shown no
relative motion since its discovery.  However, when we plot the published
speckle measurements as a function of time, both separation and position
angle exhibit a periodicity of about a decade.

\begin{figure}


\mycaption{fig.dsco-sep}{Separation and position angle of \dSco\relax}
{
Separation and position angle measurements from published interferometric
observations of \dSco\ \cite{M+H88,MHF90}, together with the MAPPIT result.
The solid curves show the orbit described in the text.  A different symbol
is used for each of the three orbital cycles covered by the data.  Open
circles indicate measurements belonging to Orbit~1 and asterisks indicate
Orbit~2.  The single measurement on Orbit~3 is from MAPPIT ($\Box$).
}

\end{figure}

\begin{figure}


\mycaption{fig.dsco-orbit2}{Orbital motion of \dSco\ with a preliminary orbit}
{
The orbital motion of \dSco\ relative to the primary~(+).  The data and
symbols are the same as for \figref{fig.dsco-sep}.  The calculated orbit is
superimposed, with tick marks showing the position at intervals of six
months during the cycle we have referred to as Orbit~2.

}
\end{figure}

 \Figref{fig.dsco-sep} shows the separation and position angle of the
\dSco\ system from catalogued speckle observations, together with the
MAPPIT result.  We see that the binary has just begun the third orbital
cycle since its discovery (or, rather, re-discovery).  The motion of the
secondary relative to the primary is shown in \figref{fig.dsco-orbit2}.

\begin{table*}
\mycaption{tab.dsco-orbit}{dummy}{
Orbital elements for \dSco}
\begin{center}
\begin{tabular}{ccccccc}
\hline\hline
$P$      & $T$     & $a$       & $e$  &   $i$  & $\omega$ & $\Omega$~(2000)\\
10.5\,yr&1979.3&0\farcs11&0.82&70\deg&170\deg&0\deg\\
\hline\hline
\end{tabular}
\end{center}
\end{table*}

The solid lines in these diagrams show an orbit with the elements given in
\tabref{tab.dsco-orbit}.  I calculated these values using
trial-and-error, combined with single-parameter least-squares fitting to
four of the elements ($T$, $P$, $e$ and $a$).  The agreement with the
observations is quite good and there is no need at this stage to refine the
fitting procedure.

Finally, we note that the orbit of \dSco\ is highly elliptical.  It is
possible that radial velocity changes will become measurable at the
turn of the century during the next periastron, which would provide
information about the masses in the system.  We can estimate the expected
radial-velocity signature as follows.

Suppose the primary and secondary have masses $M_1$ and $M_2$.  The
velocity curve of the primary star will have a semi-amplitude of
\cite{Hei}:
\begin{displaymath}
  K_1 = \frac{2\pi a' \sin i}{P\sqrt{1-e^2}(1+M_1/M_2)}.
\end{displaymath}
Here, $a'$ is the semi-major axis of the orbit, which we can estimate from
its angular size ($a=0\farcs11$) and the assumed distance of the star
(135\,pc), giving $a'=2.2\times10^9$\,km.

Based on the magnitude difference, the mass ratio for the system should be
1.5--2, depending on the spectral type of the secondary \cite{Lang}.
Combining this with the orbital elements in \tabref{tab.dsco-orbit} gives
$K_1=23$--28\,km\,s$^{-1}$.  This amplitude is 3--4 times larger than the
20~day variation reported by \citename{vHBD63} (see \secref{sec.hist}) and
so should be easily measurable during several months around periastron.  If
spectral lines from both components can be measured, both masses and the
distance could be determined directly.  The study of systems which are both
visual and spectroscopic binaries is a powerful method of determining
fundamental stellar quantities.

{\em Note added in proof.}  Cot\'e \& van Kerkwijk (A\&A, in press) have
discovered \dSco\ to be a Be star, with the H$\alpha$ line showing emission
on the flanks of an absorption core.  This makes \dSco\ one of the
brightest known Be stars (the {\em Bright Star Catalogue\/} lists only four
Be stars that are brighter).  Previous observations in the literature have
shown no indication of Be behaviour in \dSco\ and, interestingly, the
Cot\'e \& van Kerkwijk spectrum was taken only ten months after the last
periastron of the system.  It is tempting to speculate that the approach of
the companion (about 600\,R$_{\odot} \simeq 85$ stellar radii) triggered
the mechanism responsible for Be emission.  I am grateful to Rens Waters
for bringing this new result to my attention.

\subsection*{Acknowledgments}

MAPPIT is supported by a grant under the CSIRO Collaborative Program in
Information Technology.  The observations and data reduction were made in
collaboration with J.~G.~Robertson and R.~G.~Marson.  I also thank
J.~G.~Robertson for helpful comments on this manuscript.  This research has
made use of the Simbad database, operated at CDS, Strasbourg, France.

\end{document}